\documentclass[conference]{IEEEtran}

\IEEEoverridecommandlockouts

\usepackage{cite}
\usepackage{amsmath,amssymb,amsfonts}
\usepackage{algorithmic}
\usepackage{graphicx}
\usepackage{textcomp}
\usepackage{xcolor}
\usepackage{xurl}     

\definecolor{myblue}{rgb}{0.0, 0.2, 0.6} 
\definecolor{mygreen}{rgb}{0.0, 0.5, 0.0} 
\definecolor{mygray}{rgb}{0.5, 0.5, 0.5} 

\usepackage{url}

\usepackage{makecell}

\usepackage{longtable}
\usepackage{caption}

\usepackage{threeparttable} 

\usepackage[utf8]{inputenc}
\usepackage{algorithmic}
\usepackage{algorithm}
\usepackage{amsfonts} 
\usepackage{amsmath}
\usepackage{xcolor}
\usepackage{makecell}
\usepackage{dsfont}
\usepackage{mathtools}
\usepackage{subcaption}
\usepackage{tikz}

\usepackage{pifont}
\newcommand{\cmark}{\ding{51}}%
\newcommand{\xmark}{\ding{55}}%

\usepackage{colortbl} 

\newcommand{\bad}{\cellcolor{red!10}\textcolor{red!70!black}{\ding{115}}} 
\newcommand{\good}{\cellcolor{green!10}\textcolor{green}{\large$\star$}} 
\newcommand{\medium}{\cellcolor{yellow!10}\textcolor{orange}{\large$\bullet$}} 

\usepackage{booktabs}

\usepackage{float}

\newtheorem{definition}{Definition}

\newcommand{\ArchiNAME}{\texttt{Mosaic}}
\newcommand{\AlgoNAME}{\texttt{Pilot}}

\newcommand{\TxAllo}{\texttt{TxAllo}}

\newcommand{\FutureLeakageRatio}{\beta}

\newcommand{\capacity}{\lambda}

\newcommand{\ledger}{\mathcal{L}}
\newcommand{\Block}{\mathcal{B}}

\newcommand{\Tx}{\mathtt{Tx}}

\newcommand{\shard}{\mathcal{S}}
\newcommand{\shardacc}{\mathcal{A}}

\newcommand{\allocation}{\phi}

\newcommand{\numshard}{k}
\newcommand{\Difficulty}{\eta}

\newcommand{\workload}{\omega}
\newcommand{\throughput}{\Lambda}

\newcommand{\setaccount}{\mathcal{A}}
\newcommand{\setalltx}{\mathcal{T}}
\newcommand{\setallmr}{\mathcal{MR}}
\newcommand{\oneaccount}{\nu}
\newcommand{\setcross}{\mathcal{T}^\mathcal{C}}
\newcommand{\setintra}{\mathcal{T}^\mathcal{I}}
\newcommand{\workloadaverage}{\bar{\workload}}
\newcommand{\Transactionacc}{\mathcal{A}_{\Tx}}

\newcommand{\timeslot}{t}
\newcommand{\miners}{\mathcal{M}}
\newcommand{\BC}{\mathcal{BC}}
\newcommand{\MigraReq}{\mathtt{MR}}

\newcommand{\gas}{\xi}

\newcommand{\potential}{\mathcal{P}^{\oneaccount}}

\newcommand{\connection}{\psi^{\oneaccount}}
\newcommand{\WorkDis}{\Omega}

\newcommand{\ConnDis}{\Psi^{\oneaccount}}

\newcommand{\recentTauTx}{\setalltx_{[(t-\tau),t]}}
\newcommand{\recentTauMR}{\setallmr_{[(t-\tau),t]}}
\newcommand{\Txa}{\setalltx^{\oneaccount}}

\def\BibTeX{{\rm B\kern-.05em{\sc i\kern-.025em b}\kern-.08em
    T\kern-.1667em\lower.7ex\hbox{E}\kern-.125emX}}

\begin{document}
\title{\ArchiNAME: Client-driven Account Allocation Framework in Sharded Blockchains}

\author{\IEEEauthorblockN{Yuanzhe Zhang}
\IEEEauthorblockA{\textit{The University of Sydney}\\
Sydney, Australia \\
yzha9691@uni.sydney.edu.au}
\and

\IEEEauthorblockN{Shirui Pan}
\IEEEauthorblockA{\textit{Griffith University}\\
Gold Coast, Australia \\
s.pan@griffith.edu.au}
\and

\IEEEauthorblockN{Jiangshan Yu$^{\ast}$}\thanks{*Corresponding author}
\IEEEauthorblockA{\textit{The University of Sydney}\\
Sydney, Australia \\
jiangshan.yu@sydney.edu.au}
}


\maketitle

\begin{abstract}
Recent account allocation studies in sharded blockchains are typically miner-driven, requiring miners to perform global optimizations for all accounts to enhance system-wide performance. 
This forces each miner to maintain a complete copy of the entire ledger, resulting in significant storage, communication, and computation overhead.

In this work, we explore an alternative research direction by proposing \ArchiNAME{}, the first client-driven framework for distributed, lightweight \textit{local optimization}. 
Rather than relying on miners to allocate all accounts, \ArchiNAME{} enables clients to independently execute a local algorithm to determine their residing shards. 
Clients can submit migration requests to a beacon chain when relocation is necessary. 
\ArchiNAME{} naturally addresses key limitations of miner-driven approaches, including the lack of miner incentives and the significant overhead.
  While clients are flexible to adopt any algorithm for shard allocation, we design and implement a reference algorithm, \AlgoNAME{}, to guide them. 
  Clients execute \AlgoNAME{} to maximize their own benefits, such as reduced transaction fees and confirmation latency.

  On a real-world Ethereum dataset, we implement and evaluate \AlgoNAME{} against state-of-the-art miner-driven global optimization solutions. 
  The results demonstrate that \ArchiNAME{} significantly enhances computational efficiency, achieving a four-order-of-magnitude reduction in computation time, with the reduced input data size from 1.44 GB to an average of 228.66 bytes per account.
  Despite these efficiency gains, \AlgoNAME{} introduces only about a 5\% increase in the cross-shard ratio and maintains approximately 98\% of the system throughput, demonstrating a minimal trade-off in overall effectiveness.

\end{abstract}

\section{Introduction}
Blockchain technology, as a decentralized and secure ledger, holds transformative potential across various sectors, including cryptocurrency~\cite{hashemi2020cryptocurrency}, Web3~\cite{wang2022exploring}, and Decentralized Finance (DeFi)~\cite{schar2021decentralized}. 
According to CoinMarketCap, leading blockchain systems such as Bitcoin and Ethereum are valued at over US\$2.02 trillion and US\$469 billion, respectively\footnote{\url{https://coinmarketcap.com}, data retrieved on 05-Dec-2024.}.
This work focuses on permissionless, account-based blockchains like Ethereum, which allow anyone to join or leave the network freely. 
These blockchains typically involve two key participants: clients, who initiate transactions, and miners (also referred to as replicas, consensus nodes, or validators), who process transactions and maintain the ledger through consensus algorithms.

Due to the decentralized nature, scalability is a significant challenge in permissionless blockchain systems~\cite{wang2019sok, han2021security,sui2022auxchannel,sui2022monet,qin2023blindhub}.  
Blockchain sharding is considered a promising solution to address this issue.
It divides the blockchain into parallel processing shards, thereby enabling throughput to scale linearly with the number of shards.
However, cross-shard transactions, requiring expensive multi-round cross-shard consensus, continue to pose a significant challenge in the practical implementation of sharding.

Account allocation mechanisms have been recently investigated to reduce the occurrence of cross-shard transactions. Existing solutions rely on miners to execute an allocation algorithm, which can be broadly classified into two categories: `hash-based' and `graph-based' approaches. 
Hash-based approaches typically allocate accounts to shards based on the hash value of their address ID, such as SHA256(ID). 
These methods are computationally simple and provide a uniform distribution of Transaction Processing Workload (hereafter referred to as workload) across shards, optimizing the efficiency of computation, storage, communication, and workload balance. However, they fail to consider interaction dependencies between accounts, leading to over 90\% of transactions being cross-shard in sharded blockchains~\cite{wang2019sok}.

To address this issue, state-of-the-art solutions have explored graph-based allocation approaches~\cite{fynn2018challenges, mizrahi2020blockchain, huang2022brokerchain, zhang2023txallo, wang2024orbit} to further optimize the number of cross-shard transactions globally. 
These approaches require miners to use sophisticated allocation algorithms to derive an account-shard mapping, often employing techniques such as graph cutting, clustering, or community detection on a historical transaction graph. This enables shards to operate more independently, with fewer costly cross-shard operations.

While these graph-based algorithms naturally consider both the number of cross-shard transactions and the workload balance with a huge throughput boost, they introduce several inherent limitations. 
First, these approaches rely on the complete historical account interaction patterns for global optimization, which necessitates that miners synchronize and store the entire ledger (e.g., 20.1 TB in Ethereum\footnote{\url{https://etherscan.io/chartsync/chainarchive}, data retrieved on 05-Dec-2024.}). This, in turn, introduces significant computational overhead for miners with longer runtime.
Second, although miners are tasked with this resource-intensive account allocation process, the incentive for performing such tasks remains unknown. While the computational burden of these allocation algorithms may be relatively minor in proof-of-work blockchains, it introduces substantial overhead in other blockchain systems, such as those based on proof-of-stake. 
Third, miner-driven approaches require a global deterministic algorithm to avoid the need for additional consensus on allocation results. This constraint limits the flexibility to employ a wider range of probabilistic optimization techniques.

\textbf{Novel paradigm.} To overcome these open challenges, we
present \ArchiNAME{}, the first client-driven approach for account
allocation in sharded blockchains.  \ArchiNAME{} introduces a
\textit{local optimization} framework, enabling clients to
independently determine the residing shard of their accounts according to their need,
while maintaining the same level of system performance compared to
state-of-the-art solutions.

Each client is incentivized to select its residing shard based on individual preferences, such as reduced transaction fees and improved confirmation latency.
This can be achieved by minimizing cross-shard transactions and choosing a less crowded shard. 
This decision is informed by the client’s historical and expected future transaction activities.
When a client decides to change its residing shard, it submits an account migration request to the blockchain. Miners then collect and validate these submissions. Once approved, the migrations are recorded on the beacon chain (similar to the Ethereum Beacon Chain), providing all miners with a consistent view of allocation.

While clients are flexible in choosing the preferred algorithm for account allocation, we propose a reference algorithm called \AlgoNAME{} to guide them in their decision-making. It operates to optimize clients' own benefits such as transaction gas fees and conformation latency, by analyzing two key types of information: the distribution of an account's interactions across shards and the distribution of workload across shards. We define a cost function that incorporates this information, guiding clients to select shards that are less congested and more likely to facilitate a higher volume of intra-shard transactions.

We evaluate \AlgoNAME{} against state-of-the-art miner-driven global optimization solutions, including both hash-based and graph-based approaches, using over 91 million transactions from Ethereum. 
The simulation demonstrates that \AlgoNAME{} significantly reduces execution time, achieving a four-orders-of-magnitude improvement from about $0.4$ second to about $2*10^{-5}$ second. 
In our simulation, the input data size for shard allocation algorithms is reduced from 1.44 GB per miner to an average of 228.66 bytes per account. 
Notably, despite using such minimal data for local optimization, \AlgoNAME{} results in only about a 5\% increase in the cross-shard ratio while maintaining approximately 98\% of the system throughput, indicating a minimal impact on overall performance.

\begin{figure}[ht]
\centering
\includegraphics[width=0.48\textwidth]{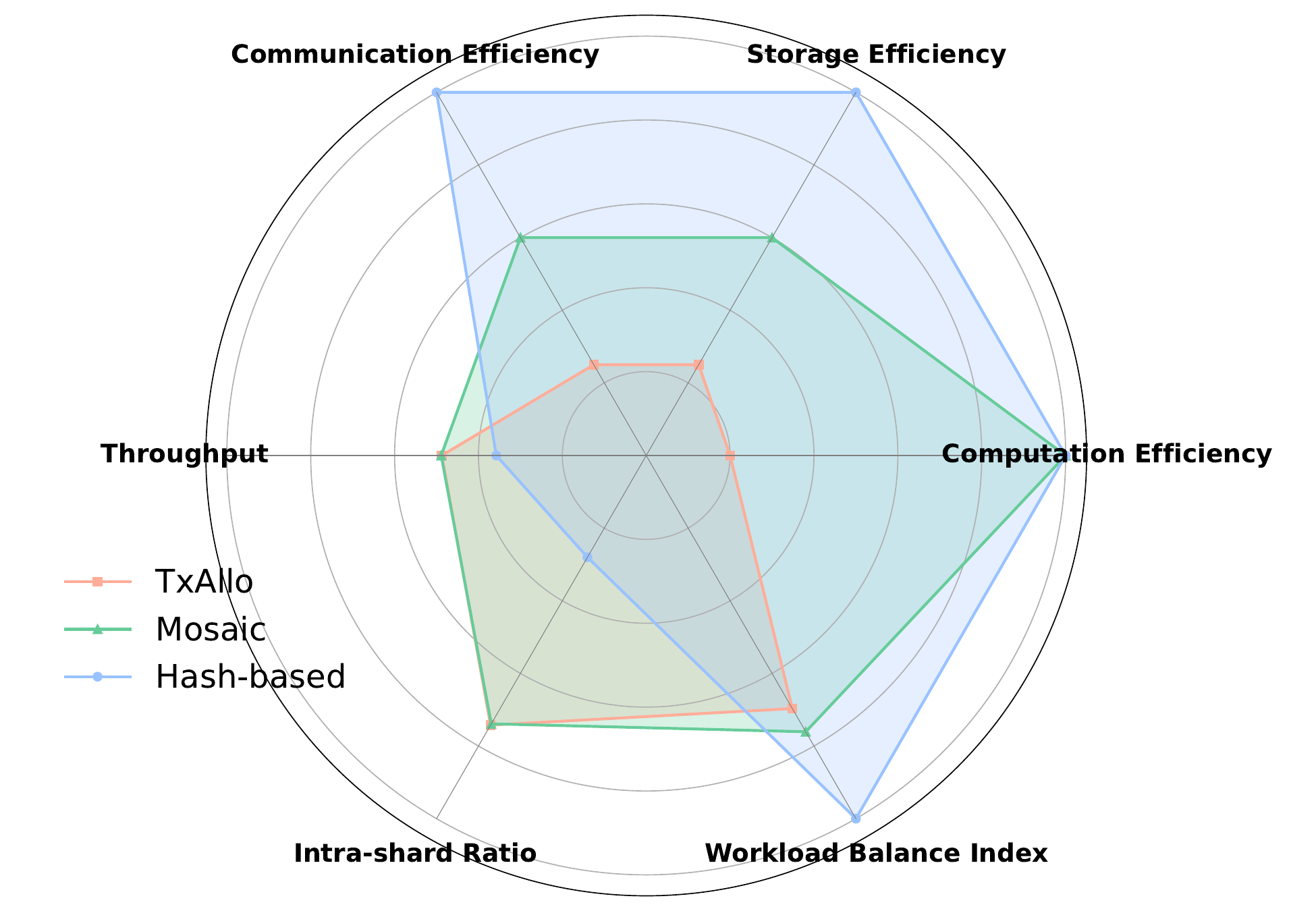}
\caption{\small \ArchiNAME{} presents a trade-off between efficiency and effectiveness compared to the baselines.}
\label{Radar_chart}
\end{figure}

Figure~\ref{Radar_chart}\footnote{The values are based on the average results from our experiments and the theoretical analysis in Section~\ref{sec_experiment} and Section~\ref{sec_comparison}, with default configuration set to 16 shards and $2.66 \times 10^{8}$ accounts as in the current Ethereum (\url{https://etherscan.io/chart/address}). To ensure comparability, the maximum and minimum values across all dimensions are normalized to 5 and 1, respectively. If a theoretical maximum exists, it is normalized to 5; otherwise, the best-performing value is normalized to 5. To maintain the principle that higher values indicate superior performance, efficiency is defined as the reciprocal of overhead, and the workload balance index is defined as the reciprocal of workload deviation.} provides an intuitive comparison, highlighting the trade-off between efficiency and effectiveness when compared to the hash-based method and the state-of-the-art graph-based method, $\TxAllo$~\cite{zhang2023txallo}.
A more detailed discussion will be provided in Section~\ref{sec_experiment} and Section~\ref{sec_comparison}.

In summary, the main contributions are as follows:
\begin{enumerate}
\item 
We propose a novel client-driven paradigm, \ArchiNAME{}, for account allocation in sharded blockchains.

\item We present \AlgoNAME{}, a reference algorithm for local optimization on client-side benefits.

\item We implement and evaluate \AlgoNAME{} under \ArchiNAME{}, highlighting significantly faster running with comparable effectiveness to baselines. 
\end{enumerate}

\section{Background and Related Work} 
\label{sec_BG}

\subsection{Blockchain Sharding} 
Blockchain sharding divides a blockchain into smaller units or `shards' for concurrent processing, in terms of both miners and transactions~\cite{han2021security}. 
Miners in each shard are responsible for processing only a subset of the total transactions allocated to this shard.
 Elastico~\cite{luu2016secure} sets a precedent that it employs a random allocation of miners to different shards, followed by a periodic reshuffling of miners during the reconfiguration phase, to mitigate the risk of single-shard takeover attacks by malicious miners. 
Subsequent permissionless sharding protocols~\cite{luu2016secure, kokoris2018omniledger, zamani2018rapidchain, wang2019monoxide} also incorporate this periodic reshuffling procedure as a security guarantee~\cite{wang2019sok}.
This ensures that malicious miners are not gathered in a specific shard for a long time.
When a miner is shuffled to a new shard, it synchronizes the state and transactions of this target shard via the blockchain's peer-to-peer network.
This enables the miner to process the upcoming transactions in the target shard in the next epoch.

In the sharding design of Ethereum 2.0\footnote{\url{https://github.com/ethereum/EIPs/blob/master/EIPS/eip-2982.md}}, a beacon chain is introduced as a separate chain, which has already been launched in 2020.  
The beacon chain is designed to support and secure the consensus of a number of parallel shard chains to fully support sharded smart contract execution in future.
The beacon chain performs operations including coordinating miners, assigning them to the shards, receiving updates from shards, processing stakes, and rebalancing shards.
This type of primary coordination chain is a common infrastructure in many systems. 
For example, the Relay Chain in Polkadot~\cite{wood2016polkadot} coordinates a network of parallel blockchains or parachains.
Similarly, the Cosmos Hub in Cosmos~\cite{kwon2019cosmos} keeps track of the state of each zone and facilitates inter-blockchain communication.
The NEAR protocol\footnote{ \url{https://nearprotocol. com/downloads/Nightshade.pdf}} implements a sharded blockchain where the main blockchain coordinates a set of shards. 
This widely used beacon chain-like architecture is also leveraged in \ArchiNAME{} to validate and store client-proposed account migration requests and to coordinate miners in synchronizing the state of the newly assigned shards.

\subsection{Transaction and Account Allocation}
Conventional sharding protocols typically use hash-based methods to allocate accounts and transactions, which are static and regardless of transaction patterns.
In particular, Chainspace~\cite{al2017chainspace} allocates accounts to shards based on the hash value of their addresses (SHA256(address) mod $k$), and transactions are processed in the shards maintaining the corresponding accounts. 
Monoxide~\cite{wang2019monoxide} similarly allocates accounts to the $2^k$-th shard by the first $k$ bits of the hash value of their addresses.
OmniLedger~\cite{kokoris2018omniledger} and RapidChain~\cite{zamani2018rapidchain}, which use the UTXO-based model(Unspent Transaction Output), both randomly allocate UTXOs based on the first several bits of their hash value, similar to Monoxide.
SharDAG~\cite{chengshardag} designed for DAG-based sharded blockchains (Directed Acyclic Graph) does not specify the rules for transaction and account allocation and allows for the integration of plug-in solutions.

Recent studies that address the account and transaction allocation problem typically require miners to execute graph-based clustering algorithms.
The backbone algorithms for these solutions can be broadly categorized into two types: the Metis method~\cite{karypis1997metis} and the $\TxAllo{}$~\cite{zhang2023txallo} method.
The Metis method, as a classical multi-level graph partition algorithm, was first utilized by Fynn et al. to obtain the partition of accounts~\cite{fynn2018challenges}. 
 Avi et al. further extended this work by considering the memory efficiency of allocation mapping~\cite{mizrahi2020blockchain}.
Huang et al. proposed BrokerChain~\cite{huang2022brokerchain} to further address the workload balance problem by introducing a special type of participants known as brokers.

Zhang et al. proposed a novel allocation algorithm, $\TxAllo{}$~\cite{zhang2023txallo}, specifically designed for sharded blockchains. It introduces a unique optimization objective that simultaneously reduces the occurrence of cross-shard transactions and ensures workload balance across shards.
Additionally, it proposes a fast algorithm to adaptively update the account-shard mapping.
Building on this work, Wang et al. further introduced Orbit~\cite{wang2024orbit} to leverage pending transactions to estimate future patterns, which are then fed into $\TxAllo{}$ as input.

However, all of these methods are miner-driven, requiring each miner to redundantly compute the allocation of all accounts based on the complete transaction graph. This approach necessitates synchronizing, storing, and processing the full ledger to execute the allocation algorithm. Although consensus is conducted independently within each shard, the associated communication and storage overhead remain substantial.

\section{\ArchiNAME}
\label{sec_design}
This section introduces a client-driven account allocation framework, which is named \ArchiNAME{} as it assembles final allocation results from many migration requests.
We assume that each client holds a single account, both of which are denoted as $\oneaccount$. 
If a client holds multiple accounts, \ArchiNAME{} follows a similar process, with the allocation decision repeated for each account individually.

\subsection{Blockchain Model}
\label{Sec: Blockchain Model}
This section formalizes the underlying blockchain model for the proposed client-driven framework \ArchiNAME{}.

\subsubsection{Sharded Blockchain}
We consider an account-based sharded blockchain as a decentralized ledger, denoted by $\ledger := (\shard_{1}, \shard_2, ..., \shard_k, \BC)$, consisting of $k$ shards and a beacon chain $\BC$. 
Each shard $\shard_i$ is composed of a chain of blocks, defined as $\shard_i := (\Block_{1}^{\shard_i}, \Block_{2}^{\shard_i}, ..., \Block_{|\shard_i|}^{\shard_i})$, where $|\shard_i|$ represents the size of the chain for the $i$-th shard. 
The beacon chain stores successfully committed account migration requests, denoted by $\BC := (\Block_{1}^{\BC}, \Block_{2}^{\BC}, ..., \Block_{|\BC|}^{\BC})$.

\begin{figure*}[t]
    \centering
\includegraphics[width=0.98\textwidth]{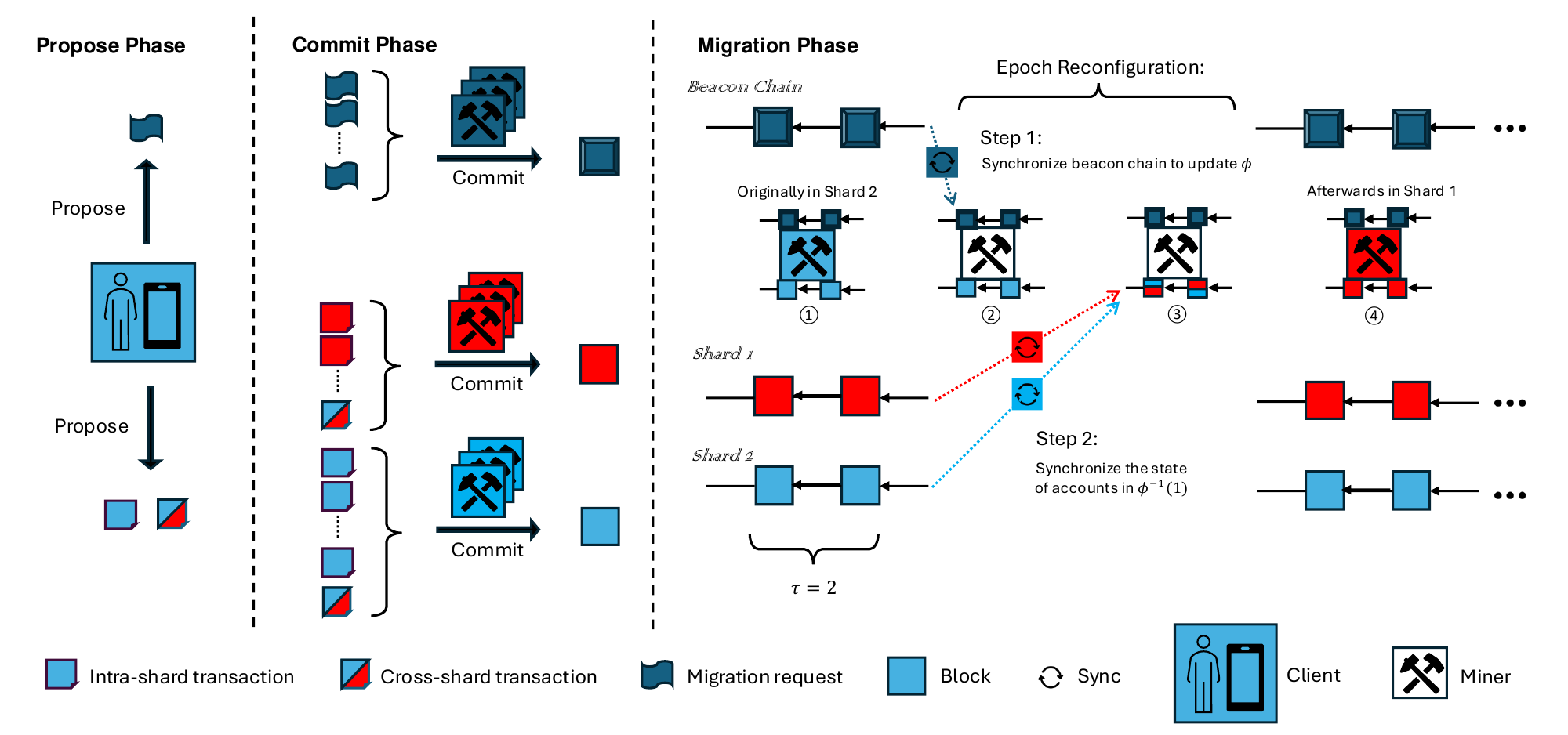}
    \caption{\small An illustration of the \ArchiNAME{} architecture with $\numshard = 2$ shards and $\tau = 2$ blocks serves as a toy example. The red and blue colours represent shard 1 and shard 2, respectively. 
During the propose phase, a client in shard 2 can propose three types of transactions: migration requests, intra-shard transactions, and inter-shard transactions. 
In the commit phase, miners in the corresponding shards or the beacon chain collect and package these transactions into a block, which is then added to the chain.
Every two blocks ($\tau = 2$), the epoch reconfiguration takes place, during which miners are reshuffled across shards and account migration occurs simultaneously. 
In this figure, a miner is reshuffled from shard 2 to shard 1 during this phase. The miner first synchronizes with the beacon chain to update its locally stored allocation mapping, $\allocation$. Then, it synchronizes with the relevant shards based on $\allocation$ to obtain the state of the accounts in the newly assigned shard 1.}
    \label{block_archi}
\end{figure*}

Let $\miners$ be the set of all miners maintaining the blockchain. 
We denote the miners for maintaining the beacon chain $\BC$ and the  shard $\shard_i$ as $\miners_{\BC}$ and $\miners_{\shard_i}$, respectively.
The beacon chain $\BC$ is accessible to all miners in $\miners$.
Let $\setalltx$ and $\setaccount$ be the set of all transactions and accounts that appeared in $\ledger$, respectively.
Let $\setallmr$ be the set of all migration requests in the beacon chain. 
Assuming that the consensus protocol running in the beacon chain is the same as the one in each local shard, the upper bound of $| \setallmr |$ is the same as the upper bound of one shard $| \setalltx |/k$. 
Every miner in $\miners$ locally stores the account-shard mapping $\allocation$.
During the account migration phase, they update $\allocation$ with the latest blocks of $\BC$.

Each shard $\shard_i$ uniquely manages a subset $\shardacc_i $ of accounts, where
$\shardacc_i \subseteq \setaccount$ and $1	\leqslant i \leqslant k$.
We formally define account allocation as an account-shard mapping:
\begin{definition}[Account-shard Mapping $\allocation$] \label{allocation_def}
    For a blockchain $\ledger$ with $k$ shards, an account-shard mapping $\allocation$  maps each account $\oneaccount \in \setaccount$ to shard ID $i\in [1, k]$. Let $\shardacc_i \subseteq \setaccount$ be the set of accounts allocated to shard $\shard_i$, i.e., $\shardacc_i = \allocation^{-1}( i)$. The tuple $\{\shardacc_1, \shardacc_2, \ldots, \shardacc_k\}$ satisfies
 \begin{itemize}
  \item Uniqueness: $\forall i, j \in [1,k]$ and $i \neq j$, $\shardacc_i \cap \shardacc_j = \varnothing$;
  \item Completeness, $\setaccount=\cup_{i=1}^k{\shardacc_i}$.
\end{itemize}
\end{definition}

A transaction $\Tx$ to be processed generally modifies the state of some particular accounts. Let the set of accounts modified by $\Tx$ be $\Transactionacc \subseteq \setaccount$. For $1 	\leqslant i \leqslant k$, if $\Transactionacc \cap \shardacc_i \neq \varnothing$, $\Tx$ will be processed in shard $\shard_i$. 
If there exists $i \neq j$ such that both $\shard_i$ and $\shard_j$ process $\Tx$, the transaction $\Tx$ is called a cross-shard transaction. Otherwise, it is an intra-shard transaction.


\subsubsection{Parameters}
To quantify the processing difficulty for shards when handling intra-shard and cross-shard transactions, we introduce the difficulty parameter $\Difficulty$. A value of 1 represents the difficulty of processing an intra-shard transaction, while $\Difficulty > 1$ reflects the increased difficulty of processing a cross-shard transaction, often due to the more complex consensus process involved. The value of $\Difficulty$ can be adjusted and may vary across different sharding protocols.

The processing capability parameter, $\lambda$, represents the maximum workload a shard can handle. We assume all shards $\shard_i$ have equal processing power due to periodic miner reshuffling, which evenly distributes computing resources across shards.
A shard can process up to $\lambda$ intra-shard transactions (difficulty = 1). However, for cross-shard transactions, where the difficulty is $\Difficulty$, a shard can handle only $\lambda / \Difficulty$ transactions.

\subsection{Architecture}
\label{Sec: Blockchain Architecture}
Figure~\ref{block_archi} illustrates the architecture of \ArchiNAME{} with two parallel shards and a single beacon chain as an example.  
To ensure a consistent view of account allocation, \ArchiNAME{} utilizes a beacon chain, which contrasts with miner-driven methodologies such as $\TxAllo{}$, requiring the allocation algorithm to be deterministic. \ArchiNAME{} consists of three logical phases: transaction proposal, transaction commitment, and account migration.

During the transaction proposal phase, clients propose two types of transactions, common transactions $\Tx$ and migration requests $\MigraReq$.
Common transactions $\Tx$ are proposed to different shards according to $\allocation$, while migration requests $\MigraReq$ are proposed to the beacon chain when a client plans to migrate its account from one shard to another.

During the transaction commitment phase, miners $\miners_{\shard_i}$ in each shard, $i=1,2,...,k$, execute a consensus protocol to commit transactions $\Tx$ and include them into the new block of $\shard_i$, according to the allocation $\allocation$. 
Simultaneously, $\miners_{\BC}$ execute a local consensus to commit migration requests $\MigraReq$ and include them into the new block of $\BC$. 
Note that this allocation $\allocation$ refers to the one from the last epoch reconfiguration, not to the current state of $\BC$, which we will discuss in Section~\ref{account_mig_phase}.

\subsubsection{Account Migration} \label{account_mig_phase}
During the account migration phase, states of accounts are moved from one shard to the other, according to the update on the beacon chain. 
Note that generally graph-based approaches and \ArchiNAME{} require account migration~\cite{huang2024account}, except for hash-based approaches relying on static allocation.

\ArchiNAME{} integrates account migration as part of the reconfiguration process at each epoch.
This reconfiguration phase is a conventional step in blockchain sharding schemes~\cite{HanYZ20, wang2019sok, luu2016secure}, where miners are systematically reshuffled across shards.
In the subsequent epoch, each miner assumes responsibility for its newly assigned shard, denoted as $\shard_j$.
In traditional reconfiguration protocols, every miner is aware of its target shard $\shard_j$ and the accounts statically assigned to $\shard_j$, represented as $\allocation^{-1}(j)$. 
At this point, miner reshuffling begins. During this phase, miners communicate with one another to synchronize the state of accounts in $\allocation^{-1}(j)$.

The key addition in \ArchiNAME{} to the conventional epoch reconfiguration process is the dynamic account allocation $\allocation$. 
In \ArchiNAME{}, the reconfiguration phase begins when miners synchronize with the beacon chain to acquire the latest blocks committed during the previous epoch.
Using these new blocks, miners update their locally stored account-shard mapping $\allocation$ to reflect the latest state in $\BC$. 
This dynamic update ensures that the current account allocation $\allocation$ is consistent across all miners.

Once a miner has updated its mapping $\allocation$ and is aware of its assigned shard $\shard_j$, it can identify the updated set of accounts $\allocation^{-1}(j)$ assigned to $\shard_j$. 
At this stage, account migration takes place concurrently with miner reshuffling. Miners synchronize the state of accounts in $\allocation^{-1}(j)$ through the same inter-shard communication process used for reshuffling.

The dynamic account migration introduced in \ArchiNAME{} does not introduce additional communication overhead. This is because both account migration and miner reshuffling share the same synchronization phase.
 Specifically, account migration follows the updated account-shard mapping $\allocation$, while miner reshuffling adheres to the rules established by existing epoch reconfiguration protocols.

The epoch reconfiguration phase occurs at fixed intervals, every $\tau$ blocks on the beacon chain, which defines the epoch length. Even if block generation rates differ across individual shards and the beacon chain, the reconfiguration phase still takes place as scheduled. The process remains consistent regardless of the number of blocks generated in each shard during the epoch.

\subsubsection{Overhead of Account Migration}
The design of performing account migration during the epoch reconfiguration phase aims to minimize synchronization overhead for miners. Synchronizing beacon chain and shard information at arbitrary intervals can lead to inconsistencies across different miners' states. Therefore, synchronization must occur at a pre-defined pace. By setting this pace to $\tau$ blocks during epoch reconfiguration, \ArchiNAME{} avoids additional communication and storage overhead.

\ArchiNAME{} does not modify existing miner reshuffling protocols.
Miners update the state of accounts in their newly assigned shards following the same process as in conventional sharding protocols, maintaining an equivalent volume of data transfer and storage.
While reducing the synchronization interval to fewer than $\tau$ blocks would result in a more current account allocation $\allocation$, it would also introduce additional synchronization phases and overhead.

By leveraging the epoch reconfiguration phase, the only additional overhead introduced by account migration is the storage of migration-related transactions on the beacon chain. 
These migration requests are processed as local transactions on the beacon chain, and their number is limited by the capacity of the consensus algorithm. Consequently, the overhead is minimal compared to existing miner-driven approaches, which often require significant computation, storage, and communication resources.

\subsection{Decision-making Information}
This section introduces two crucial factors for clients to make decisions in effective shard selection: a client $\oneaccount$'s Interaction Distribution $\ConnDis$ across shards and the current Workload Distribution $\WorkDis$ across shards.

\subsubsection{Interaction Distribution}
Interaction distribution describes the frequency of an account's transactions with different shards, providing insights into potential cross-shard transactions. 
If an account frequently interacts with others in a particular shard, it benefits the client to choose that shard to minimize cross-shard transactions. 
The interaction distribution is maintained by each account $\oneaccount$ as a local vector $\ConnDis = [\connection_1, \connection_2,...,\connection_k] \in \mathbb{R}^{k} $, which is a $k$-dimensional vector. 
 Each entry $\connection_i$ in this vector indicates the number of interactions that $\oneaccount$ has with a specific shard $\shard_i$.
In this way, the vector $\ConnDis$ serves as an estimation of the interactions of the account with all shards.

The information on the interaction distribution is derived from two main sources. 
The first source is the blockchain's committed transaction records.
  Once a transaction is fully committed, it provides publicly accessible information on the interacted accounts and their respective shards. 
  Each client can store these committed interactions related to himself for future analysis. 
  The historical committed transactions for an account, $\oneaccount$, are referred to as $\Txa_h$.

For a transaction $\Tx$, let $\setaccount_{\Tx}$ be the set of accounts associated with $\Tx$. 
 $\ConnDis_h = [\connection_{h,1}, \connection_{h,2},...,\connection_{h,k}]$ is computed by the equation:
\begin{equation}\label{equ_trans2conn}
    \connection_{h,i} = \sum_{\Tx \in \Txa_h}\sum_{\mathtt{b} \in \setaccount_{\Tx}-\{\oneaccount\}}\mathds{1}(\allocation(\mathtt{b})=i).
\end{equation}
 $\mathds{1}()$
is the indicator function, which outputs 1 if
$\allocation(\mathtt{b})=i$ and 0 otherwise.
This equation indicates that $ \connection_{h,i}$ counts the times of that $\oneaccount$ interacts with other accounts $\mathtt{b} \in \allocation^{-1}(i)$ in transactions in $\Txa_h$.
$\setaccount_{\Tx}-\{\oneaccount\}$ indicates the set difference operation on $\setaccount_{\Tx}$ and $\{\oneaccount\}$.

  The second source of information for interaction distribution is the client’s own knowledge of their expected transactions, influenced by factors such as daily routines, geographical location, and social connections. 
  For example, daily activities and local ties may involve transactions with specific service providers or businesses, while social relationships could indicate future transactions with friends or family. 
  These expected transactions are denoted as $\Txa_e$.
  We can compute $\connection_e$ using $\Txa_e$, based on Equation~(\ref{equ_trans2conn}).

We define a parameter, $\FutureLeakageRatio$, as the ratio of known expected future transactions to all future transactions, which also reflects the client’s confidence in their estimated future transaction patterns. The account  $\oneaccount$ can combine these two sources of information into a final fused interaction distribution, $\connection$, for comprehensive analysis. 
A standard method for this combination is outlined in Section~\ref{sec_algorithm}.

\subsubsection{Workload Distribution}
The workload distribution across various shards is the other crucial information for clients to minimize the latency of their transaction confirmation.
By choosing a shard with a lower workload, clients can ensure faster transaction processing, thereby enhancing their overall experience.

In permissionless blockchains, clients can obtain workload distribution information through public service platforms. 
Taking Ethereum as an example, Etherscan\footnote{ \url{https://etherscan.io/txsPending}} is a well-known platform that analyzes the transaction mempool (a repository of pending transactions) to forecast gas prices\footnote{ \url{https://etherscan.io/gastracker}}. 
This makes the workload distribution of the entire network openly accessible to all participants, allowing any entity to easily observe and acquire this information. 
In future deployments of \ArchiNAME{}, similar functionality could be extended to make workload distributions of different shards publicly available, enabling clients to simply communicate with these platforms and download a minimal amount of data, eliminating the need for on-chain storage.

Similar to $\ConnDis$, each client $\oneaccount$ locally maintains a vector to record the workload distribution across shards, $\WorkDis = [\workload_1, \workload_2, ..., \workload_k] \in \mathbb{R}^{k} $. 
 Each entry $\workload_i $ of $\WorkDis$ indicates the current workload in a specific shard $\shard_i$, obtained from public platforms.
The communication and storage overhead required for clients to update the vector $\WorkDis$ is negligible as $\WorkDis$ comprises a minimal amount of data, only $k$ numbers.

\section{\AlgoNAME}
\label{sec_algorithm}
This section introduces \AlgoNAME{}, a streamlined algorithm that provides standardized shard selection guidance, offering clients a practical tool within the \ArchiNAME{} framework to facilitate easy engagement.
It is important to note that \AlgoNAME{} serves as a guiding tool to benefit clients, but it does not mandate its suggestions, allowing clients the flexibility to choose their shard and corresponding allocation policy.

\textbf{Future Knowledge Fusion.} Let $\FutureLeakageRatio$ be the ratio of known expected future transactions out of all future transactions.
The fused transaction pattern $\ConnDis$ from the historical and expected pattern is computed by: 
\begin{equation}
\label{Equ_fusion}
\ConnDis = (1-\FutureLeakageRatio) * \ConnDis_h + \FutureLeakageRatio * \ConnDis_e.
\end{equation}

The equation implies that a client $\oneaccount$ would rely more on the expected future transactions if $\oneaccount$ has more confidence in knowing more future transactions.
In the extreme case, when $\oneaccount$ knows all future transactions,  $\oneaccount$  can fully rely on the expected patterns that $\ConnDis = \ConnDis_e$.
In the worst case,  when $\oneaccount$ does not know any future transactions,  $\oneaccount$  can only rely on the historical patterns that $\ConnDis = \ConnDis_h$.

\textbf{Cost Function.} The cost function is defined for client-side benefits but it matches the desired system-side benefits.
It is defined as the total cost for processing the transactions of $\oneaccount$, if $\oneaccount$ is in shard $\shard_i$:
\begin{equation}
\label{Equ_cost}
    u_{i}^{\oneaccount} = (1*\connection_{i}+\Difficulty*\connection_{-i})*\gas_{i} + \Difficulty\sum_{j\neq i}{\connection_{j}*\gas_{j}},
\end{equation}
where $\connection_{-i} = \sum_{j \neq i}{\connection_{j}}$ represents the sum of interactions in all other shards except shard $\shard_i$.

The notation $\gas_{i}$ is the standard cost for a transaction being processed in shard $\shard_i$.
Taking Ethereum as an example, $\gas_{i}$ can stand for the gas fee (tip or priority fee) paid to miners or burned.
If a shard $\shard_i$ is more overloaded, i.e., with higher $\workload_i$, $\oneaccount$ needs to pay more $\gas_{i}$ to make $\oneaccount$' transactions being processed successfully.
Thus, $\gas_{i}$ is the result of a monotonic increasing function $f$ of $\workload_i$, $\gas_i = f(\workload_i)$.
For simplicity, we let $\gas_i = f(\workload_i) = \workload_i$ in \AlgoNAME.
One can design a more specialized function $f$ for the specific needs of applications.

The first term in Equation~\ref{Equ_cost}, $1*\connection_{i}*\gas{i}$, represents the total cost of intra-shard transactions within shard $\shard_i$. 
For cross-shard transactions involving $\oneaccount$, where transactions must be processed in both $\shard_i$ and other shards $\shard_j$, the costs are captured by the other two additional terms. 
The second term, $\Difficulty*\connection_{-i}*\gas_{i}$, accounts for the part of cross-shard processing costs in $\shard_i$. 
The third term, $\Difficulty*\sum_{j\neq i}{\connection_{j}*\gas_{j}}$, covers the costs of processing the remaining parts of cross-shard transactions in other shards $\shard_j$.

\textbf{Simplified Computing.}
We introduce an easy way to compute $u_{i}^{\oneaccount}$.
Let $\connection :=  \sum_{j \in \{1,2,...,k\}}{\connection_{j}}$.
Suppose that $\oneaccount$ is comparing the total cost when selecting $\shard_i$ and $\shard_j$.
\begin{equation*}
\begin{aligned}
    u_{i}^{\oneaccount}-u_{j}^{\oneaccount} & = (1*\connection_i+\eta*\connection_{-i}) * \workload_{i} + \eta\sum_{p\neq i}{\connection_{p}*\workload_{p}} \\
    & - (1*\connection_{j}+\eta*\connection_{-j})*\workload_{j} - \eta\sum_{q\neq j}{\connection_{q}*\workload_{q}} \\
     =  & [1*\connection_i+\eta*(\connection - \connection_{i})] * \workload_{i} + \eta\connection_j\workload_j \\
    & - [1*\connection_{j}+\eta*(\connection - \connection_{j})]*\workload_{j} - \eta\connection_i\workload_i \\
     = & (2\Difficulty-1)\connection_j\workload_j-(2\Difficulty-1)\connection_i\workload_i+\Difficulty\connection\workload_i-\Difficulty\connection\workload_j \\
     = & [(2\Difficulty-1)\connection_j-\Difficulty\connection]*\workload_j -[(2\Difficulty-1)\connection_i-\Difficulty\connection]*\workload_i
\end{aligned}
\end{equation*}

The equation above indicates that $u_{i}^{\oneaccount}-u_{j}^{\oneaccount} < 0$, if and only if $[(2\Difficulty-1)\connection_j-\Difficulty\connection]*\workload_j -[(2\Difficulty-1)\connection_i-\Difficulty\connection]*\workload_i < 0$.
Thus, we denote it as a new concept \textit{Potential}:
\begin{equation}\label{equ_potnetial}
    \potential_i = [(2\Difficulty-1)\connection_i-\Difficulty\connection]*\workload_i.
\end{equation}
From the above analysis, the shard $\shard_i$ with a minimum of $u_{i}^{\oneaccount}$ is the one with a maximum of $\potential_i$.
Comparing with $u_{i}^{\oneaccount}$, $\potential_i$ only computes $\connection_i$ and $\workload_i$ of shard $\shard_i$ rather than  $\connection_j$ and $\workload_j$  in all other shards.

Although $u_{i}^{\oneaccount}$ is designed from the perspective of clients to reduce their costs, it aligns well with system-level metrics, cross-shard ratio and workload balance.
This is achieved by encouraging a high number of interactions $\connection_i$ and a low workload $\workload_i$.
We analyze this through $\potential_i$.

From the perspective of reducing the cross-shard ratio, as the weight of $\connection_i$ is positive, i.e., $(2\Difficulty -1)*\workload_i > 0$,
clients tend to reside in the shard $\shard_i$ with higher $\connection_i$ to achieve higher potential $\potential_i$.

From the perspective of selecting a shard with a low workload, there are two possible cases regarding the weight $[(2\Difficulty-1)\connection_i-\Difficulty\connection]$ that can be positive or negative. 
If there exists a shard $\shard_i$ where this weight is positive, we have $\connection_i/\connection > \Difficulty/(2\Difficulty-1) > 1/2$.
This indicates that most interactions of $\oneaccount$ are with shard $\shard_i$. 
Consequently, for any other shard $\shard_j$ (where $j \neq i$), the weight $[(2\Difficulty-1)\connection_j-\Difficulty\connection] < 0$ and thus, $\potential_i > 0 > \potential_j$. 
This implies that if $\oneaccount$ is highly connected to a shard $\shard_i$, Equation~\ref{equ_potnetial} encourages $\oneaccount$ to reside in $\shard_i$, regardless of the workload $\workload_i$.
In the other case where the weight is negative for all shards $\shard_i$, $\oneaccount$ is encouraged to reside in the shard $\shard_i$ with the least $\workload_i$ to maximize $\potential_i$.

\textbf{Pseudo Algorithm.} The pseudo code for \AlgoNAME{} is presented in Algorithm~\ref{cdrec}. 
The goal of the algorithm is to determine the shard allocation mapping $\allocation(\oneaccount)$ for a specific account $\oneaccount$.
The algorithm seeks to identify the shard $\shard_i$ that offers the highest potential $\potential_i$ for the account $\oneaccount$.
Lines 1-4 compute the fused connection distribution $\connection$ based on historical and expected transactions, which is a key factor in calculating $\potential_i$ in Equation~\ref{equ_potnetial}.
Lines 5-14 iterate over all shards to find the shard with the highest potential $\potential_i$. 
The shard with the highest potential corresponds to the shard that minimizes the cost $u_{i}^{\oneaccount}$ for account $\oneaccount$.

\begin{algorithm}[htbp]
 \begin{normalsize}
 \caption{\AlgoNAME{}}
 \label{cdrec}
 \begin{algorithmic}[1]
\REQUIRE
    \textcolor{myblue}{$\Txa_{h}$}: historical committed transactions for account \textcolor{myblue}{$\oneaccount$};  
    \textcolor{myblue}{$\Txa_{e}$}: expected transactions for account \textcolor{myblue}{$\oneaccount$};  
    \textcolor{myblue}{$\WorkDis$}: workload distribution;  
    \textcolor{myblue}{$\allocation$}: current allocation; 
    \textcolor{myblue}{$\numshard$}: number of shards; 
    \textcolor{myblue}{$\Difficulty$}: difficulty parameter for processing cross-shard transactions
\ENSURE
    \textcolor{myblue}{$\allocation(\oneaccount)$}: updated allocation for \textcolor{myblue}{$\oneaccount$}

\STATE \textcolor{mygreen}{// Compute historical and expected connection distributions}
\STATE Compute \textcolor{myblue}{$\ConnDis_h$} and \textcolor{myblue}{$\ConnDis_e$} using Equation~(\ref{equ_trans2conn}).

\STATE \textcolor{mygreen}{// Fuse the historical and expected distributions}
\STATE Compute \textcolor{myblue}{$\ConnDis$} using Equation~(\ref{Equ_fusion}).

\STATE \textcolor{mygreen}{// Calculate the potential of the current allocation}
\STATE Compute \textcolor{myblue}{$\potential_{\allocation(\oneaccount)}$} using Equation~(\ref{equ_potnetial}).

\FOR{\textcolor{mygray}{$i \in \{1, 2, ..., \numshard\} \setminus \{\allocation(\oneaccount)\}$}}
    \STATE \textcolor{mygreen}{// Calculate the potential of shard $\shard_i$}
    \STATE Compute \textcolor{myblue}{$\potential_{i}$} using Equation~(\ref{equ_potnetial}).
    \IF{\textcolor{mygray}{$\potential_{i} > \potential_{\allocation(\oneaccount)}$}}
        \STATE \textbf{Update the allocation:} \textcolor{myblue}{$\allocation(\oneaccount) = i$}
        \STATE \textbf{Update the potential:} \textcolor{myblue}{$\potential_{\allocation(\oneaccount)} = \potential_{i}$}
    \ENDIF
\ENDFOR
\RETURN \textcolor{myblue}{$\allocation(\oneaccount)$}

  \end{algorithmic}
  \end{normalsize}
\end{algorithm}

\section{Evaluation} \label{sec_experiment}
Given the complexities in building a real distributed system with clients' behaviours, this section presents a simulation based on real-world Ethereum data. 
The results show that \AlgoNAME{} under the framework \ArchiNAME{} achieves comparable performance in terms of cross-shard transaction ratio and workload balance across shards and throughput, while the execution time is negligible compared to prior algorithms.

\subsection{Experiment Setup}
\textbf{Environment.}
The experimental simulations were implemented using Python 3.8, implemented on a computing cluster node equipped with an Intel Xeon Gold 6150 CPU and 55 GB of memory.

\textbf{Dataset.}
As a leading platform, Ethereum has a long history of exploring sharded blockchain structures and maintains an extensive repository of real-world transactions. 
Although we use Ethereum data for our experiments, it is important to note that \ArchiNAME{} can be seamlessly applied to other sharded blockchain systems as well.

We collected a dataset using Ethereum ETL\footnote{\url{https://ethereum-etl.readthedocs.io/en/latest/google-bigquery/}}, covering transactions from block 10,000,000 to block 10,600,000. 
This dataset spans approximately three months and includes over 91 million transactions and 12 million accounts.
The first 90\% of the dataset is used for the initial allocation, while the remaining 10\% is reserved for evaluation. 
The experiments are conducted over 200 epochs, with $\tau = 300$ blocks per epoch, which corresponds to approximately one hour in the Ethereum network.
Evaluation metrics are calculated using the data from the current epoch based on the allocation results computed at the end of the preceding epoch.

\textbf{Parameters.}
We conduct experiments using different values for the difficulty parameter, denoted as $\Difficulty$ (2, 5, and 10), and the number of shards, $\numshard$ (4, 16, and 32). The default experimental setting is $\Difficulty = 2$ and $\numshard = 16$. 
To avoid extremely overloaded or underloaded cases, we set $\capacity = |\setalltx^{[(\timeslot-\tau),\timeslot]}|/\numshard$ transactions per epoch for every shard.
By this configuration, in the best case where all transactions are intra-shard and evenly distributed across shards, the overall throughput is exactly $|\setalltx^{[(\timeslot-\tau),\timeslot]}|$ transactions per epoch.

Since \ArchiNAME{} operates with a single beacon chain, we set the maximum number of committed $\MigraReq$ (migration requests) in one epoch to $\capacity$, which is bounded by the processing capacity. If the number of proposed $\MigraReq$s exceeds this limit, the migration requests that offer the most significant improvements in $\potential$ will be prioritized for commitment.

In \ArchiNAME{}, workload distribution $\WorkDis$ is from open platforms analyzing the mempool of pending transactions. 
Thus, it is from analyzing transactions in the next epoch in this simulation.
The workload $\workload_i$ of $\shard_i$ is set as the total workload to process transactions in it, $\workload_i= |\setintra_i|+ |\setcross_i|*\Difficulty$, where $\setintra_i$ and $\setcross_i$ are the set of intra-shard and cross-shard transactions in $\shard_i$, respectively.

As clients simultaneously consider migrating their residing shards, we set $\allocation(\setaccount_{\Tx}-\{\oneaccount\})$ to the current allocation. While clients may have the potential to communicate, form alliances, or compete, the development of a more sophisticated simulated model is out of the scope of this paper.

\textbf{Evaluation metrics and baselines.}
The evaluation of effectiveness includes the metrics below, showing that \ArchiNAME{} achieves comparable performance with graph-based miner-driven methods.
\begin{itemize}
    \item \textbf{Cross-Shard Transaction Ratio} is the ratio of cross-shard transactions to the total number of transactions. A lower ratio is preferable for improving performance.

\item \textbf{Workload Deviation} quantifies the balance of the workload across shards, calculated as the standard deviation of the workload across shards, $\big(\frac{\sum_{i=1}^{k} (\workload_i-\workloadaverage)^{2}} {k\workloadaverage} \big)^{0.5}$. 
Here, $\workload_i$ represents the workload of shard $i$, and $\workloadaverage$ is the average workload across all shards. A lower deviation indicates a more balanced workload across the shards.

\item \textbf{System Throughput} indicates the total number of transactions that the system can handle in a unit of time, which follows the same metrics as in $\TxAllo$~\cite{zhang2023txallo}. Higher throughput implies a more efficient system.
\end{itemize}

The evaluation of efficiency includes the metrics below, showing that \ArchiNAME{} achieves scale-level efficiency boost with graph-based miner-driven methods.
\begin{itemize}
\item \textbf{Execution Time}  is the time it takes for a client to execute \AlgoNAME. A shorter execution time indicates a more efficient allocation algorithm.
\item \textbf{Input Data Size}  is the storage size for a client to execute \AlgoNAME. A smaller input data size indicates a more efficient algorithm in storage.

\end{itemize}

The baselines include the following.
\begin{itemize}
\item \textbf{Hash-based random allocation}~\cite{al2017chainspace, wang2019monoxide, kokoris2018omniledger, zamani2018rapidchain} allocates accounts by the hash value of their address IDs, which can be viewed as random allocation to the transaction patterns. 
\item \textbf{Graph-based allocation using Metis}~\cite{fynn2018challenges,mizrahi2020blockchain, huang2022brokerchain} allocates accounts by a classical graph partition algorithm Metis.
\item \textbf{Graph-based allocation using \TxAllo{}}~\cite{zhang2023txallo, wang2024orbit} is specifically designed for sharded blockchains as a state-of-the-art graph-based allocation. It includes a complete algorithm $\texttt{G}-\TxAllo{}$ on the full ledger and a fast algorithm $\texttt{A}-\TxAllo{}$ on the recent ledger. 
\end{itemize}

\subsection{Effectiveness Comparison}
\label{Sec_performance}
This section compares effectiveness metrics, including cross-shard ratio, throughput, and workload deviation, shown in Tables~\ref{table_ratio}, \ref{table_throughput}, and \ref{table_wbi} with different parameters.
 In the first time step, the allocation $\allocation$ is initialized by the result of $\TxAllo$ for $\AlgoNAME$. 
For new accounts that do not appear in the historical data, $\TxAllo$ and Metis-based methods are not capable of assigning them. Therefore, these accounts are randomly allocated.
To demonstrate the performance superiority of $\AlgoNAME$ even in its worst case, our experiment considers the scenario where clients have no knowledge about future transactions, i.e., $\FutureLeakageRatio = 0$.

\begin{table}[!htb]
\caption{The comparison of average cross-shard ratios. The bold figures denote the optimal values. Values in parentheses indicate the performance loss relative to the best baseline, which has an average value of 5.06$\%$.}
\centering
\begin{tabular}{lcccc}
\toprule
Parameters & \textbf{\AlgoNAME{}} & $\TxAllo{}$ & Metis & Random\\
\midrule
 k = 4 & 24.07$\%$ (5.20$\%$) & 25.64$\%$ & \textbf{22.88}$\%$ & 74.95$\%$ \\ 
 k = 16 & \textbf{33.58$\%$} (0$\%$) & 35.94$\%$ & 33.92$\%$ & 92.96$\%$  \\ 
 k = 32 & 42.24$\%$ (9.89$\%$) & 39.21$\%$ & \textbf{38.44$\%$} & 96.03$\%$ \\ 
$\Difficulty$ = 5 & 31.55$\%$ (4.99$\%$) & \textbf{30.05}$\%$ & 33.90$\%$ & 92.96$\%$  \\ 
$\Difficulty$ = 10 & 31.78$\%$ (5.23$\%$) &  \textbf{30.20}$\%$ & 33.90$\%$ & 92.96$\%$  \\ 
\bottomrule
\end{tabular}
\label{table_ratio}
\vspace{-0.5em}
\end{table}

\subsubsection{Comparison of Cross-shard Transaction Ratio}
\label{sec_cross_ratio}
Table~\ref{table_ratio} provides a comprehensive comparison of the cross-shard transaction ratios. 
\AlgoNAME{} effectively reduces the number of cross-shard transactions, achieving values comparable to miner-driven methods such as $\TxAllo$ and Metis. 
On average, the cross-shard transaction ratio is only 5.06$ \%$ higher than the best miner-driven baseline. 
The hash-based random allocation performs the worst, as it fails to utilize transaction patterns.

The $5.06\%$ performance loss is from various factors. Negative factors include: \AlgoNAME{} does not leverage the full graph as input but only the account-related transactions, and clients have no knowledge about others' allocations. The positive factor is that new accounts of the system can allocate themselves using \AlgoNAME{}.

\subsubsection{Comparison of Throughput}
\label{sec_throughput}
We employ a normalized throughput, ${\throughput}/{\capacity}$, to clearly illustrate how many times the throughput is improved by sharding. 
This normalization sets the throughput of a non-sharded blockchain (i.e., when $k=1$) as the benchmark, scaled to 1.
Remarkably, as shown in Table~\ref{table_throughput}, \AlgoNAME{} presents on average only 1.85$\%$ performance loss to the best-performing baseline.

\begin{table}[!htb]
\caption{The comparison of average throughput improvement. The bold figures denote the optimal values. Values in parentheses indicate the performance loss relative to the best baseline, which has an average value of 1.85$\%$.}
\centering
\begin{tabular}{lcccc}
\toprule
Parameters & \textbf{\AlgoNAME{}} & $\TxAllo{}$ & Metis & Random\\
\midrule
 k = 4 & \textbf{2.34} (0$\%$) & 2.23 & 2.31  & 1.22 \\ 
 k = 16 & \textbf{7.60} (0$\%$) & 7.31 & 7.21  & 4.21\\ 
 k = 32 & 13.06 (4.18$\%$) & \textbf{13.63} & 13.26 & 8.23 \\ 
$\Difficulty$ = 5 & 3.69 (1.60$\%$) & \textbf{3.75} & 3.68  & 1.70 \\ 
$\Difficulty$ = 10 & 1.95 (3.47$\%$) & 1.95 & \textbf{2.02}  & 0.85 \\
\bottomrule
\end{tabular}
\label{table_throughput}
\end{table}

\subsubsection{Comparison of Workload Balance}
\label{sec_wlb}
Achieving a balanced workload for account-shard mapping is challenging due to inherently imbalanced transaction patterns.
Table~\ref{table_wbi} provides a comprehensive comparison of the workload deviation to evaluate workload balance. 
Remarkably, \AlgoNAME{} outperforms the miner-driven baselines, $\TxAllo$ and Metis. 
This is mainly because the workload information is derived from the mempool, which encodes future information, and new accounts can allocate themselves by the workload distribution information.
The hash-based random allocation performs the best, as it naturally distributes accounts in an even way.

\begin{table}[!htb]
\caption{The comparison of average workload deviation. The bold figures denote the optimal values. Values in parentheses indicate the performance loss relative to the best baseline, which has an average value of 47.35$\%$.}
\centering
\begin{tabular}{lcccc}
\toprule
Parameters & \textbf{\AlgoNAME{}} & $\TxAllo{}$ & Metis & Random\\
\midrule
 k = 4 & 0.22 (29.41$\%$)& 0.22 & 0.30 & \textbf{0.17} \\ 
 k = 16 & 0.59 (51.28$\%$) & 0.73 & 0.71 & \textbf{0.39} \\ 
 k = 32 & 0.83 (40.67$\%$)  & 0.98 & 1.02 & \textbf{0.59}\\ 
$\Difficulty$ = 5 & 0.64 (64.10$\%$)  & 0.78 & 0.65 & \textbf{0.39}\\ 
$\Difficulty$ = 10 & 0.59 (51.28$\%$) & 0.79 & 0.63 & \textbf{0.39}\\ 
\bottomrule 
\end{tabular}
\label{table_wbi}
\end{table}

\subsection{Efficiency Comparison}
\label{sec_efficiency}
\subsubsection{Execution Time Comparison}
Table~\ref{table_running time} presents a comparison of the average execution times for different values of $k$ and $\Difficulty$. \AlgoNAME{} is significantly faster than $\TxAllo$ and Metis, demonstrating superior efficiency at scales of $10^4$ and $10^6$, respectively. The remarkable efficiency of \AlgoNAME{} comes from its innovative design. By using only a subset of relevant transactions, it avoids processing the entire transaction history, unlike $\TxAllo{}$ and Metis, which analyze the full ledger for account allocation. Additionally, a client using \AlgoNAME{} computes only their own allocation rather than the allocation for all accounts.

\subsubsection{Input Data Size Comparison}
The last row of Table~\ref{table_running time} presents a comparison of the average input data size in the simulation environment.
As \AlgoNAME{} requests clients to store only the transactions related to them, $\Txa$, the storage size for input data is dramatically smaller than baselines.
Even when compared with $\texttt{A}-\TxAllo{}$, which is the most efficient component of $\TxAllo{}$ utilizing only recent transactions for updating allocation, \AlgoNAME{} also achieves improvement at the scale of $10^3$.
This is mainly because of the individual optimization design.

\begin{table}[!htb]
\vspace{-0.5em}
\caption{The comparison of average running time  (seconds) and input data size. As $\TxAllo$ consists of two components $\texttt{A}-\TxAllo{}$ and $\texttt{G}-\TxAllo{}$, the values at the left and right-hand side of `$\backslash$' represent the results of them, respectively.}
\centering
\begin{tabular}{lccc}
\toprule
Parameters & \textbf{\AlgoNAME{}} & $\TxAllo{}$ & Metis \\
\midrule
 k = 4 & \textbf{$2.03*10^{-5}$} & 0.42 $\backslash$ 61.31& 346.58  \\ 
 k = 16 & \textbf{$2.33*10^{-5}$} & 0.38 $\backslash$ 57.98 & 387.43   \\ 
 k = 32 & \textbf{$2.61*10^{-5}$} & 0.51 $\backslash$ 55.75 & 421.74   \\ 
$\Difficulty$ = 5 & \textbf{$2.37*10^{-5}$} & 0.48 $\backslash$ 40.70 & 404.40   \\ 
$\Difficulty$ = 10 & \textbf{$2.49*10^{-5}$} & 0.44 $\backslash$ 80.84 & 411.75   \\ 
\cmidrule(lr){1-4}
Input Data & 228.66 Bytes & \makecell{ 721.14 KBs\\ $\backslash$ 1.44 GBs } & 1.44 GBs  \\ 
\bottomrule 
\end{tabular}
\label{table_running time}
\vspace{-1em}
\end{table}

\subsection{Impact of Future Knowledge}
This section explores the impact of the side benefit, expected future transactions. We configure all clients to share the same future knowledge parameter, $\FutureLeakageRatio$, with values set to 0, 0.25, 0.5, 0.75, and 1.
In scenarios with fewer shards, account allocation tends to be more stable because clients are more likely to retain the same allocation strategy from the previous epoch. Therefore, we set $\Difficulty = 2$ and $\numshard = 4$ as the default parameters for this analysis.
As shown in Table~\ref{table_future}, the worst performance is observed when $\FutureLeakageRatio = 0$, where clients lack any knowledge of future transactions. As $\FutureLeakageRatio$ increases, generally, performance improves as expected, since greater knowledge of future transactions allows clients to make more informed and strategic allocation decisions.
It is important to note, as highlighted in Section~\ref{Sec_performance}, that \AlgoNAME{} still achieves comparable performance even when $\FutureLeakageRatio = 0$. This demonstrates that future transaction knowledge, while beneficial for further improvement, is exploitable but not mandatory for the effectiveness of our method.

\begin{table}[!htb]
\caption{The performance comparison with various $\FutureLeakageRatio$, including cross-shard ratio, throughput improvement, workload deviation.}
\centering
\begin{tabular}{lccc}
\toprule
Metrics &  Ratio & Throughput  & Workload \\
\midrule
 $\FutureLeakageRatio$ = 0 & 24.06$\%$ & 2.34 & 0.22 \\ 
 $\FutureLeakageRatio$ = 0.25 & 20.91$\%$  & 2.45 & 0.17 \\ 
 $\FutureLeakageRatio$ = 0.5 & 20.09$\%$ & 2.48 & 0.15\\ 
$\FutureLeakageRatio$ = 0.75 & \textbf{19.73$\%$}  & \textbf{2.50} & 0.12\\ 
$\FutureLeakageRatio$ = 1 & 20.69$\%$ & 2.45 & \textbf{0.12}\\ 
\bottomrule 
\end{tabular}
\label{table_future}
\end{table}

\section{Comparison of frameworks}
\label{sec_comparison}

\begin{table*}[t]
  \caption{Comprehensive comparison. \ArchiNAME{} demonstrates comparable performance to graph-based methods while maintaining good efficiency and offering additional intrinsic benefits.}
\centering
\begin{tabular}{lcccc}
\toprule
& Graph-based methods & \ArchiNAME{} & Hash-based methods \\
\midrule
\textbf{Computation} & & & \\
Participants & Miners & Clients & Miners \\
Optimization type & Global optimization & Local optimization & Global optimization \\
Redundant computation results & $\allocation (\setaccount)$ \bad & $\allocation (\oneaccount)$ \good & $\allocation (\setaccount)$ \bad \\
Computation complexity (w.r.t. input size) & $O(|\setalltx|)$ \bad & $O(|\Txa|)$ $\ddagger$ \good & $O(|\recentTauTx|)$ \good \\
\midrule
\multicolumn{1}{l}{\textbf{Storage and Communication}} & & & \\
Replication storage & $|\setalltx|$ \bad & $|\setalltx|/k + |\setallmr|$ $^*$ \medium & $|\setalltx| / k$ \good \\
Replication communication & $|\recentTauTx|$ \bad & $|\recentTauTx|/k + |\recentTauMR|$ $^*$\medium & $|\recentTauTx|/k$ \good \\
\midrule
\textbf{Performance} & & & \\
Cross-shard ratio & \good & \good & \bad \\
Throughput & \good & \good & \bad \\
Workload balance & \medium & \medium & \good \\
\midrule
\textbf{Side Benefits} & & & \\
Computation incentives & \xmark & \cmark & \xmark \\
Allocation controllability & \xmark & \cmark & \xmark \\
Allocation of new accounts & \xmark & \cmark & \cmark \\
Future expected transactions & \xmark & \cmark & \xmark \\
\bottomrule
\end{tabular}
\begin{tablenotes}
    \footnotesize
    \item A green checkmark (\good) denotes good; an orange line (\medium) denotes medium; and a red triangle (\bad) denotes poor. A `\cmark' symbol indicates the property is achieved, while a `\xmark' symbol indicates it is not. The performance evaluation is based on experimental results from Section~\ref{sec_experiment}.
    \item 
    $\ddagger$ $|\Txa| \ll |\setalltx|$, where $|\Txa|$ represents the transactions related to a single account $\oneaccount$, on average at the scale of $2|\setalltx|/|\setaccount|$.
    \item 
    $^*$ This part focuses on miner-side replicated storage and communication required to maintain the sharding system. On the client side, \ArchiNAME{} requires clients to store only their related transactions, which is a common feature of existing wallets. 
    The client-side communication is also negligible, including submitting migration requests to the beacon chain and retrieving workload distribution data from third-party platforms.
\end{tablenotes}
\label{table_benefits_compare}
\end{table*}

This section provides a comprehensive comparison
between our proposed approach and existing miner-driven approaches, as detailed in Table~\ref{table_benefits_compare}.

\textbf{Intrinsic computation differences.} Graph-based methods require all miners to
periodically run global optimization algorithms based on the accumulative historical transactions, to derive the allocation results of all accounts $\allocation(\setaccount)$. 
Hash-based methods similarly compute redundant allocation results $\allocation(\setaccount)$.
However, only new accounts or transactions need to be processed, as the allocation for existing accounts remains static. 
Additionally, the underlying hash function, which only takes the address ID as input, is significantly faster than graph-based algorithms, making hash-based methods highly efficient.

In contrast, \ArchiNAME{} relies on clients to run the allocation
algorithms individually according to their needs and demands, based
only on their local transactions. This allows each client to run any
efficient allocation algorithm to determine the allocation of one account $\allocation(\oneaccount)$ with minimal input $\Txa$ (transactions local to the client
rather than all transactions on the entire blockchain).

\textbf{Storage and communication overhead.} Miners in graph-based
methods have to store the full historical transaction data and
periodically synchronize and update recent transaction data, with
storage and communication volumes of $|\setalltx|$ and
$|\recentTauTx|$ respectively. 
In contrast, since hash-based methods use only the address IDs as inputs, miners only need to maintain the transactions within their corresponding shards, rather than the full ledger.

In \ArchiNAME{}, miners are not required to run the allocation algorithm, so they also do not need to store the entire ledger. Their additional storage and communication overhead comes only from the beacon chain. Thus, the storage and communication requirements for miners are $|\setalltx|/k + |\setallmr|$ and $|\recentTauTx|/k + |\recentTauMR|$, respectively. Since the beacon chain follows the same consensus protocols as the shards, the upper bound of $|\setallmr|$ is approximately $O(|\setalltx|/k)$. This results in up to $2/k$ of the storage and communication overhead compared to miner-driven methods.
From the clients' perspective, \ArchiNAME{} requires them to store only the transactions directly related to them, denoted as $\Txa$.
As normally one transaction connects two accounts, on average, $O(\Txa)=O(2|\setalltx|/|\setaccount|)$.

\textbf{Side benefits.} Through its novel client-driven design, \ArchiNAME{} offers additional advantages beyond just effectiveness and efficiency. 
We discuss some of these benefits that deserve detailed further research, including incentives for allocation computation, enhanced client control over algorithms and shard selection, the ability to manage new accounts, and sources of expected future activities.

Firstly, clients in \ArchiNAME{} have natural incentives to conduct allocation algorithms to optimize their experience, such as reducing fees and improving confirmation latency.
In contrast, heavy computation of the allocation algorithm in miner-driven methods negatively impacts miners' already intensive mining tasks, without offering sufficient incentives for miners to perform such computation.

Secondly, clients have full control over both the optimization algorithms and the resulting shard allocation. 
This prevents them from being forced into unwanted shards by miner-driven methods, which often compromise individual preferences for global performance. 
Furthermore, clients can select the optimization algorithm that best fits their specific needs, allowing them to leverage advancements in optimization techniques, rather than being limited to a single deterministic algorithm.

Thirdly, graph-based methods struggle to allocate newly created accounts, as these accounts are not included in the historical transaction graph. 
In contrast, \ArchiNAME{} enables new participants to independently determine their shard allocation by proposing migration requests to the beacon chain based on their planned activities. 
This better fulfils their specific needs while also enhancing overall system performance.

Lastly, clients often have more accurate knowledge of their future transactions than miners, based on factors such as daily routines, geographical location, and social connections. 
This enables them to make more informed decisions about shard allocation. 
Since the benefits of allocation (such as lower gas fees, reduced latency, and fewer cross-shard transactions) are realized in future transactions, knowing future patterns is far more valuable than relying on historical estimates, as graph-based methods do.

\section{Discussions}
\label{sec_discussion}

\subsection{Privacy}
\ArchiNAME{} enables clients to use future knowledge for shard allocation, which may raise potential privacy concerns about future transaction leakage. 
However, this only reveals the shards a client is likely to interact with, not the specific accounts. 
Given the large number of accounts in any shard, it remains challenging to identify the exact interacting parties in a client’s future transactions.
Clients even concerned about shard interaction privacy can opt out of this feature, as it is entirely optional and based on individual preferences. Moreover, as demonstrated in Sections~\ref{Sec_performance} and \ref{sec_efficiency}, \ArchiNAME{} performs effectively even when future knowledge is not used (i.e., $\beta = 0$).

\subsection{Security}
Our design preserves the fundamental security properties of existing sharded blockchain architectures, with the only modification that allows clients to determine their residing shard. 
As a result, the overall security level remains consistent with that of traditional sharded blockchains.
The primary security concern arises from the possibility of denial-of-service (DoS) attacks, where clients could flood the beacon chain with excessive migration requests or overwhelm a single shard with excessive transactions.

However, we argue that conducting such attacks is economically irrational. 
In the case of a DoS attack on the beacon chain, its security level is comparable to that of a single blockchain. If clients were to submit a large volume of migration requests, they would incur proportional fees, making it financially unsustainable to maintain the attack. This economic mechanism is how blockchains inherently prevent DoS attacks.

A similar argument applies to DoS attacks aimed at flooding a single shard. 
Simply migrating a large number of accounts into the same shard has no detrimental effect on the shard's performance from any perspective. 
To effectively execute a DoS attack on a shard, an attacker would need to generate a high volume of transactions, which does not require a large number of accounts (even a single account can achieve this). 
However, this type of attack is inherently mitigated by the existing blockchain design, as the associated higher transaction fees make it economically unfeasible.

\subsection{Client behaviours}
In practice, clients may behave differently according to their own strategy. For example, clients may coordinate with each other for shard allocation, which would be reflected in the $\allocation(\setaccount_{\Tx}-\{\oneaccount\})$ of Equation (\ref{equ_trans2conn}). This introduces the potential for collaborated clients with enhanced performance and further optimization. However, this is a complex multi-party game where clients may misbehave or have an alliance with trusted partners. Future work on developing a systematic game-theoretical model to better understand the actual benefit of such coordination is desired.

\section{Conclusions} \label{sec_conclu}
This paper presents the first client-driven framework for account allocation in sharded blockchains, \ArchiNAME{}.
It adopts local optimization to naturally reduce the storage, communication and computation overhead and brings several intrinsic side benefits.
This paper further introduces a reference algorithm, \AlgoNAME{}, to assist clients in this shard selection process by optimizing their benefits.
Our experiments show that this approach is not only faster by four orders of magnitude but also retains nearly 95$\%$ of the system performance of state-of-the-art miner-driven methods.
As this is a pioneering work exploring client-driven methods, a lot of future research is desired, especially the comprehensive game theoretical modeling of user behaviors.

\bibliographystyle{IEEEtran}
\bibliography{MyRefs}

\end{document}